\documentclass[journal,twoculomn]{IEEEtran}
\usepackage{cite}
\ifCLASSINFOpdf
 \usepackage[pdftex]{graphicx}

\else

  \usepackage[dvips]{graphicx}

\fi

\usepackage{bm}
\usepackage{epsfig}
\usepackage{amsmath,amssymb}
\usepackage{amsthm}
\usepackage{algorithmic}
\usepackage{relsize}
\usepackage{array}

\ifCLASSOPTIONcompsoc
 \usepackage[caption=false,font=normalsize,labelfont=sf,textfont=sf]{subfig}
\else
 \usepackage[caption=false,font=footnotesize]{subfig}
 \fi
\usepackage{amssymb}
\DeclareMathAlphabet\mathbfcal{OMS}{cmsy}{b}{n}
\usepackage{color}

\usepackage{amssymb}

\begin{document}
\title{Reconfigurable Antenna Multiple Access for 5G mmWave Systems}

\author{
    \IEEEauthorblockN{Mojtaba Ahmadi Almasi\IEEEauthorrefmark{1}, Hani Mehrpouyan\IEEEauthorrefmark{1}, David Matolak\IEEEauthorrefmark{7}, Cunhua Pan\IEEEauthorrefmark{2}, Maged Elkashlan\IEEEauthorrefmark{2}}\\
    \IEEEauthorblockA{\IEEEauthorrefmark{1}{\small{Department of Electrical and Computer Engineering, Boise State University,
   \{mojtabaahmadialm,hanimehrpouyan\}@boisestate.edu}}}\\
    
    \IEEEauthorblockA{\IEEEauthorrefmark{7}{\small{Department of Electrical Engineering, University of South Carolina,
     \{matolak\}@cec.sc.edu}}}\\

 \IEEEauthorblockA{\IEEEauthorrefmark{2}{\small{School of Electronic Engineering and Computer Science, Queen Mary University of London, \{c.pan,maged.elkashlan\}@qmul.ac.uk}}}
\thanks{This project is supported in part by the NSF ERAS grant award numbers 1642865.}
\vspace{-15pt}
}


\markboth{}%
{Shell \MakeLowercase{\textit{et al.}}: Bare Demo of IEEEtran.cls for IEEE Journals}
\maketitle
\thispagestyle{empty}
\begin{abstract}
This paper aims to realize a new multiple access technique based on recently proposed millimeter-wave reconfigurable antenna architectures. To this end, first we show that integration of the existing reconfigurable antenna systems with the well-known non-orthogonal multiple access (NOMA) technique causes a significant degradation in sum rate due to the inevitable power division in reconfigurable antennas. To circumvent this fundamental limit, a new multiple access technique is proposed. The technique which is called reconfigurable antenna multiple access (RAMA) transmits only each user's intended signal at the same time/frequency/code, which makes RAMA an inter-user interference-free technique. Two different cases are considered, i.e.,  RAMA with partial and full channel state information (CSI). In the first case, CSI is not required and only the direction of arrival for a specific user is used. Our analytical results indicate that with partial CSI and for symmetric channels, RAMA outperforms NOMA in terms of sum rate. Further, the analytical result indicates that RAMA for asymmetric channels achieves better sum rate than NOMA when less power is assigned to users that experience better channel quality. In the second case, RAMA with full CSI allocates optimal power to each user which leads to higher achievable rates compared to NOMA for both symmetric and asymmetric channels. The numerical computations demonstrate the analytical findings. 
\end{abstract}
\section{Introduction}
The rapid growth of global mobile data traffic is expected to be satisfied by exploiting a plethora of new technologies, deemed as $5$th generation (5G) networks. To meet this demand, millimeter-wave (mmWave) communications operating in the $30-300$ GHz range is emerging as one of the most promising solutions~\cite{r1}. The existence of a large communication bandwidth at mmWave frequencies represents the potential for significant throughput gains. Indeed, the shorter wavelengths at the mmWave band allow for the deployment of a large number of antenna elements in a small area, which enables mmWave systems to potentially support more higher degrees of beamforming gain and multiplexing~\cite{r1}. However, significant path loss, channel sparsity, and hardware limitations are major obstacles for the deployment of mmWave systems. In order to address these obstacles, several mmWave systems have been proposed to date~\cite{r5,el2014spatially,r9}. 

An analog beamforming mmWave system is designed in~\cite{r5} which uses one radio frequency (RF) chain and can support only one data stream. Subsequent work considers a hybrid beamforming mmWave system to transmit multiple streams by exploiting several RF chains~\cite{el2014spatially}. In~\cite{r9}, the concept of beamspace multi-input multi-output (MIMO) is introduced where several RF chains are connected to a lens antenna array via switches. In addition to these systems, non-orthogonal multiple access (NOMA) has been also considered as another promising enabling technique for 5G to enhance spectral efficiency in multi-user scenarios~\cite{higuchi2015non,dai2015non}. Unlike orthogonal multiple access (OMA) techniques that are realized in time, frequency, or code domain, NOMA is realized in the power domain~\cite{tse2005fundamentals}. In fact, NOMA performs superposition coding (SC) in the power domain at the transmitter. Subsequently, successive interference cancellation (SIC) is applied at the receiver~\cite{saito2013system,saito2013non}. 

In order to serve more users in 5G wireless communications, recently, the integration of NOMA in mmWave systems, i.e., mmWave-NOMA, has been studied~\cite{ding2017random,ding2017noma,wang2017spectrum,hao2017energy,xiao2017joint}. The work in~\cite{ding2017random} designs a random beamforming technique for mmWave-NOMA systems. The base station (BS) randomly radiates a directional beam toward paired users. In~\cite{ding2017noma}, it is shown that mismatch between the users' channel vector and finite resolution analog beamforming\footnote{Finite resolution analog beamforming is due to the use of a finite number of phase shifters.} simplifies utilizing NOMA in mmWave-MIMO systems. 
The work in~\cite{wang2017spectrum}, proposes the combination of beamspace MIMO and NOMA, to ensure more users can be served with a limited number of RF chains. As a result, the number of served users is more than the number of RF chains~\cite{wang2017spectrum}. 
In~\cite{hao2017energy}, NOMA is studied for hybrid mmWave-MIMO systems. A power allocation algorithm has been provided in order to maximize energy efficiency. Newly, a joint power allocation and beamforming algorithm for NOMA in the analog mmWave systems has been proposed in~\cite{xiao2017joint}. 
Although the works~\cite{ding2017random,ding2017noma,wang2017spectrum,hao2017energy,xiao2017joint} have all resulted in maximizing bandwidth efficiency, this gain has come at the costs of higher complexity at the receiver and the use of multiple RF chains or one RF chain along with a large number of phase shifters and power amplifiers which can be costly at mmWave frequencies. Hence, in contrast to the prior works, here, we will propose a new multiple access scheme that takes advantage of reconfigurable antennas to outperform NOMA, while at the same requiring one RF chain at the transmitter and simple receiver structure.



Recently, there has been a new class of reconfigurable antennas that can support the transmission of multiple radiation beams with one RF chain~\cite{r19,schoenlinner2002wide}. The unique property that distinguishes the system in~\cite{r19,schoenlinner2002wide} from prior art is that the proposed reconfigurable antenna architecture can support multiple simultaneous orthogonal reconfigurable beams via one RF chain. Inspired by this class of antennas, this paper proposes a fresh multiple access technique for mmWave reconfigurable antenna systems which is called reconfigurable antenna multiple access (RAMA). We consider a scenario in which a single BS is equipped with a mmWave reconfigurable antenna and each beam of the antenna serves one user where the users are not aligned with the same direction. Given that the limitation on the RF circuitry of the antenna~\cite{r19} results in the division of the transmitted power amongst the beams, the current state-of-the-art in mmWave-NOMA would not operate efficiently in such a setting. To enhance the performance of multiple access schemes in the mmWave band and also overcome this fundamental limit for reconfigurable antennas, unlike NOMA, RAMA aims to transmit only the intended signal of each user. To accommodate this technique, we will consider two cases, RAMA with partial channel state information (CSI) and RAMA with full CSI. In the first case, channel gain information is not required and only the direction of arrival (DoA) for a specific user is used. Our results indicate that with partial CSI and for symmetric channels, RAMA outperforms NOMA in terms of sum rate. Further, the analytical result indicates that RAMA for asymmetric channels achieves better sum rate than NOMA when less power is assigned to a user that experiences a better channel quality. In the second case, RAMA with full CSI allocates optimal power to each user which leads to higher achievable rates compared to NOMA for both symmetric and asymmetric channels. Our extensive numerical computations demonstrate the analytical findings.


\textbf{Notations:} Hereafter, $j = \sqrt{-1}$. Also, $\mathbb{E}[\cdot]$ and $|\cdot|$ denote the
expected value and amplitude value of $(\cdot)$, respectively.
\section{System Model and NOMA}
In this section, first we introduce the reconfigurable antenna systems and their properties. Then, NOMA technique for a BS and multiple users is described. Finally, NOMA for the reconfigurable antennas is investigated. 
\subsection{Reconfigurable antenna systems}
\begin{figure}
\includegraphics[scale = 0.43]{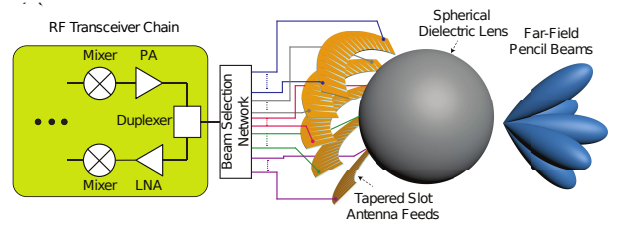}
\centering
\caption{Schematic of the reconfigurable antenna steering multiple beams. The antenna is composed of a spherical lens fed with a number of
tapered slot antenna feeds. Each feed generates a beam in a given direction in the far field~\cite{r19}.}
 \label{fig1}
\end{figure}

A reconfigurable antenna can support multiple reconfigurable orthogonal radiation beams. To accomplish this, a spherical dielectric lens is fed with multiple tapered slot antennas (TSAs), as shown in Fig.~\ref{fig1}. The combination of each TSA feed and the lens produces highly directive beams in far field~\cite{schoenlinner2002wide,r19}. That is, each TSA feed generates a beam in a given direction in the far field. Therefore, a reconfigurable antenna system is a multi-beam antenna capable of generating $M \gg 1$ independent beams where $M$ is the number of TSA feeds. Only the feed antennas that generate the beams in the desired directions need to be excited. To this end, the output of the RF chain is connected to a beam selection network (see Fig.~\ref{fig1}). The network has one input port that is connected to the RF chain and $M$ output ports that are connected to the $M$ TSA feeds \cite{r19}. For instance, in Fig.~\ref{fig1}, the network selects only five outputs that are connected to the input ports of five TSA feeds. Accordingly, the network divides power of the output signal of the RF chain equally or unequally amongst five TSA feeds. 

It is mentioned that the reconfigurable antennas steer reconfigurable independent beams. This steering is achieved by selecting the appropriate TSA feed. Also, recall that we assume that the transmitter has knowledge of the DoA of the users. Accordingly, by appropriately steering the beams, the reconfigurable antenna can manipulate the phase of the received signal at the user terminal. Therefore, steering multiple reconfigurable independent beams and routing the power amongst those beams are two properties of the reconfigurable antennas. 
\subsection{Review of NOMA}
In this paper, we consider the downlink of a single communication cell with a BS in the center that serves multiple users. The BS and users  are provided with a signal omnidirectional antennas. For simplicity, the number of users is restricted to two where the users are not aligned with the same direction, i.e., there is an angle gap between the users. 

Let the BS have signals $s_i$ ($i = 1,~2$) for User $i$, where $\mathbb{E}[|s_i|^2] = 1$, with power transmission $p_i$. The sum of $p_i$s, for $i = 1,~2$, equals to $p$. According to the principle of the NOMA downlink, at the transmitter, $s_1$ and $s_2$ are superposition coded as 
\begin{align}\label{eq1}
x = \sqrt{p_1}s_1 + \sqrt{p_2}s_2.
\end{align} 
Hence, the received signal at the $i$th user, for $i = 1,~2$, is given by
\begin{align}\label{eq22}
y_i = xh_i + n_i \equiv (\sqrt{p_1}s_1 + \sqrt{p_2}s_2)h_i + n_i,
\end{align}
where $h_i$ is the complex channel gain between the BS and User $i$, and $n_i$ denotes the additive white Gaussian noise with power $\sigma_i^2$. At the receiver, each user performs the SIC process to decode the desired signal. The optimal decoding order depends on the channel gain. Without loss of generality, let us assume that User $1$ have better channel gain, i.e., $|h_1|^2/\sigma_1^2 \geq |h_2|^2/\sigma_2^2$, which gives $p_1 \leq p_2$. 

After applying SIC, the achievable rate for NOMA for User $i$ can be determined as
\begin{align}\label{eq33}
\begin{cases}
R^N_1 = \text{log}_2(1 + \frac{p_1|h_1|^2}{\sigma^2_1}), \\
R^N_2 = \text{log}_2(1 + \frac{p_2|h_2|^2/\sigma^2_2}{p_1|h_2|^2/\sigma^2_2 + 1}).
\end{cases}
\end{align}
This result indicates that power allocation greatly affects the achievable rate for each user. For example, an improper power allocation does now allow User $1$ to decode $s_2$ correctly, which in turn does not allow for the interference from User $2$ to be successfully eliminated. 
\subsection{NOMA for the reconfigurable antennas}
Suppose that a BS is equipped with the described reconfigurable antenna system and aims to simultaneously serve two users by using NOMA. The reconfigurable antenna steers two beams by feeding two TSAs. Each TSA serves one user which is equipped with a single omnidirectional antenna. The superposition coding of $s_i$ with allocated power $p_i$ is defined in (\ref{eq1}). Users 1 and 2 receive the following signals as
\begin{align}\label{eq2}
\begin{cases}
z_1 = {\sqrt{\alpha}} xh_1 + n_1  \equiv {\sqrt{\alpha}}(\sqrt{p_1}s_1 + \sqrt{p_2}s_2)h_1 + n_1,\\
z_2 = {\sqrt{1-\alpha}} xh_2 + n_2  \equiv {\sqrt{1-\alpha}}(\sqrt{p_1}s_1 + \sqrt{p_2}s_2)h_2 + n_2,
\end{cases}
\end{align}
respectively, where factor $\alpha \in (0,1)$ is due to the power division in the reconfigurable antennas. When $\alpha = 0.5$, it means the power is divided equally between two TSAs. 
Thus, each user receives only a portion of the power of the transmitted signal $x$. 

Consider the same order for the channel gains, i.e, $|h_1|^2/\sigma_1^2 \geq |h_2|^2\sigma_2^2$. An error-free SIC process results in the achievable rate for each user as  
\begin{align}\label{eq3}
\begin{cases}
R^{N_R}_1 = \text{log}_2(1 + \frac{\alpha p_1|h_1|^2}{\sigma^2_1}), \\
R^{N_R}_2 = \text{log}_2(1 + \frac{(1-\alpha) p_2|h_2|^2/\sigma_2^2}{(1-\alpha)p_1|h_2|^2/\sigma_2^2 + 1}). 
\end{cases}
\end{align}
Obviously, signal to noise ratio (SNR) for User $1$ and signal to interference plus noise ratio (SINR) for User $2$ in (\ref{eq3}) is less than that of the NOMA given in (\ref{eq33}). As a result, combining NOMA with reconfigurable antennas will reduce the achievable user rate when considering the same channel gains, power allocation, and noise power. It is noteworthy that for reconfigurable antenna-NOMA the definition of power allocation is different from power division. Power allocation is a strategy which is used in NOMA. While, power division is one of the properties of the reconfigurable antennas that divides power of the superposition coded signal among two TSAs. 

 In brief, when users are not aligned in the same direction, the reconfigurable antennas cannot harvest the benefits of NOMA due to the power division property. Whereas, when users are located on the same direction, the reconfigurable antenna steers only one beam to serve users which means power division is not required. In this case, reconfigurable antenna-NOMA and NOMA in Subsection II.B achieve identical sum rate performance.    
\section{Reconfigurable Antenna Multiple Access}
In this section, a novel multiple access technique for the reconfigurable antenna systems is proposed. The technique which we call RAMA takes advantage of the reconfigurable antenna and directional transmission in mmWave bands. 

\subsection{RAMA with Partial CSI}
Let us assume that the BS utilizes a reconfigurable antenna and has partial CSI, i.e., knows the DoA of users. Thus, power is allocated equally for each user, i.e., 
\begin{align} \label{eq3a}
p_1 = p_2 = 0.5p. 
\end{align}
 
 Our main objective is to suppress inter-user interference. To this end, we aim to transmit only the intended signal for each user at the same time/frequency/code blocks. The intended signals for Users $1$ and 2 are $s_1$ and $s_2$, respectively. Let us assume that $s_i$, for $i = 1,~2$, is drawn from a phase shift keying (PSK) constellation and $\mathbb{E}[|s_i|^2] = 1$. Accordingly, $s_2$ can be expressed in terms of $s_1$ as 
\begin{align}\label{eq5}
s_2 = s_1e^{j\Delta \theta},
\end{align}
where $\Delta\theta$ denotes the difference between the phases of $s_1$ and $s_2$. Similar to (\ref{eq1}), the power of the transmitted signal, $x$, is assumed to be $p$. 

Unlike NOMA, in RAMA only one of the signals, say $s_1$, is upconverted by the RF chain block and the whole power $p$ is allocated to that signal before the power division step. Therefore, $x$ is given by  
\begin{align}\label{eq4}
x = \sqrt{p}s_1.
\end{align} 
It is clear that the superposition coded signal in (\ref{eq1}) and the signal in (\ref{eq4}) carry the same average power $p$. 
\begin{figure}
\includegraphics[scale = 0.25]{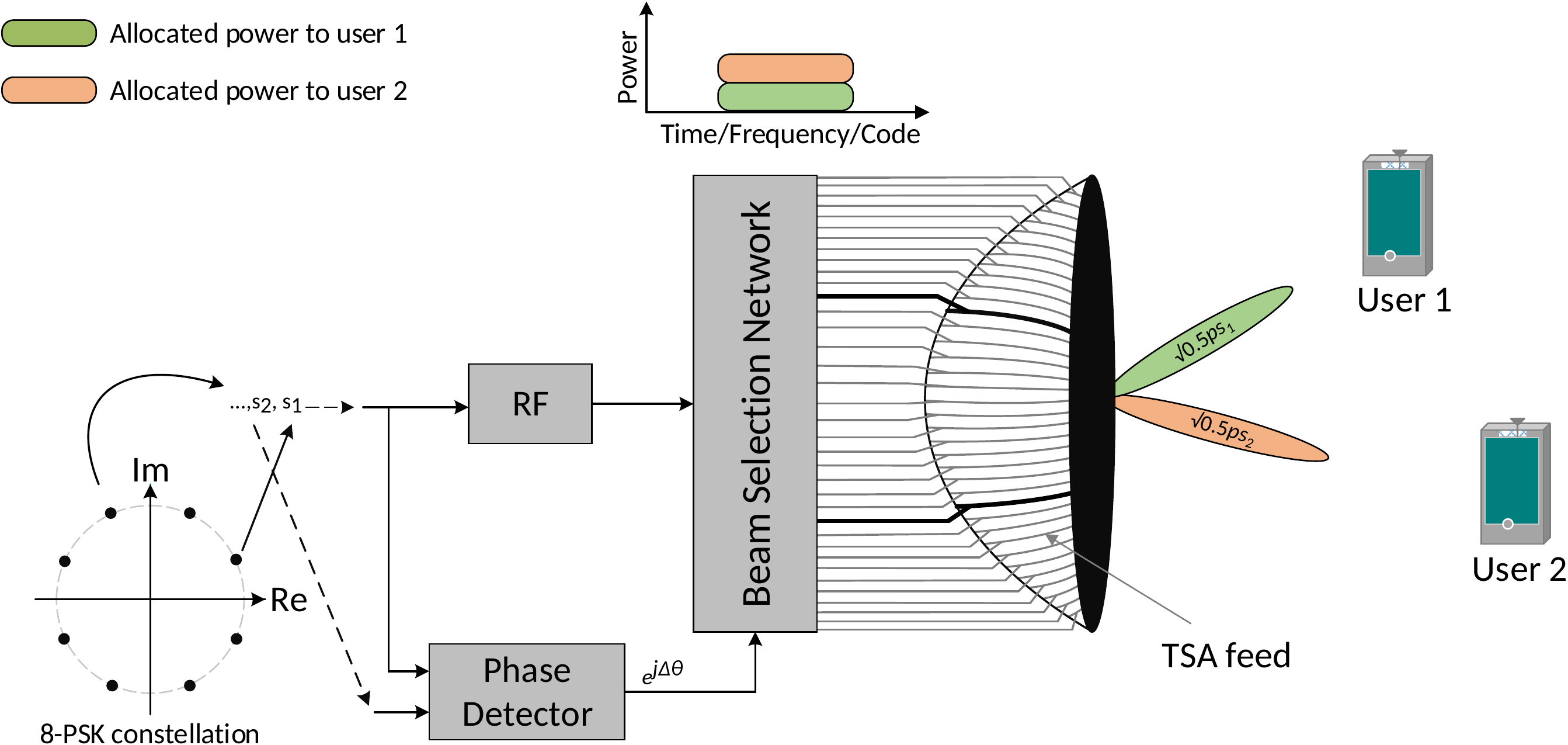}
\centering
\caption{Schematic of the BS for reconfigurable antenna multiple access technique with partial CSI and equal power division amongst the TSA feeds. It is assumed that the signals $s_1$ and $s_2$ are selected form $8$-PSK constellation}
 \label{fig2}
\end{figure}
The proposed multiple access technique for the signal in (\ref{eq4}) is shown in Fig.~\ref{fig2}. The phase detector block calculates the phase difference between $s_1$ and $s_2$, $e^{j\Delta\theta}$. Moreover, as shown in Fig.~\ref{fig2}, the beam selection network selects two TSA feeds as highlighted with black color based on the users' DoA. For simplicity, we call them TSA 1 and TSA 2. The network divides the power equally between TSA 1 and TSA 2. That is, the signal of TSAs 1 and 2 is given by $\sqrt{0.5p}s_1$.

The signal intended for TSA 1 is the desired signal for User 1. To transmit the signal intended for User 2 via TSA 2 we take advantage of the reconfigurable antennas. Thanks to properties of reconfigurable antennas, signal at each TSA can independently be rotated with an arbitrary angle. Using this proper, the beam selection network multiplies the signal in corresponding to TSA 2 with $e^{j\Delta\theta}$. This results in the transmitted signal via TSA 2 to be $\sqrt{0.5p}s_1e^{j\Delta\theta} = \sqrt{0.5p}s_2$ which is the desired signal for User 2. Since transmission is directional in mmWave bands, each user receives only its intended signal as    
\begin{align}\label{eq6}
\begin{cases}
z_1 = \sqrt{{0.5p}}s_1h_1 + n_1,\\
z_2 = \sqrt{{0.5p}}s_2h_2 + n_2.\\
\end{cases}
\end{align}
It is noteworthy that, here, we assume that there is no interference that is imposed from the signal intended for User 1 on User 2 and vice versa. This assumption is very well justified since the structure of the proposed lens based slotted reconfigurable antenna results in very directional beams with very limited sidelobes. Moreover, due to significant pathloss and shadowing at mmWave frequencies we do not expect the signals from the sidelobes to reach the unintended user\footnote{Detail analysis of the impact of sidelobes on inter-user interference in RAMA is subject of future research.}. We also highlight that in contrast to NOMA, full CSI and the SIC process are not required at the receiver. Furthermore, in RAMA power allocation and power division carry the same concept, such that the routed power for TSA 1 (or 2) is the same as the allocated power for User 1 (or 2).     

The achievable rate of RAMA for each user under equal power allocation is obtained as 
\begin{align}\label{eq7}
\begin{cases}
R^{R,I}_1 = \text{log}_2(1 + \frac{p|h_1|^2}{2\sigma^2_1}), \\
R^{R,I}_2 = \text{log}_2(1 + \frac{p|h_2|^2}{2\sigma^2_2}). 
\end{cases}
\end{align}

Let us denote RAMA with partial CSI by RAMA-I. It is valuable to compare the sum rate of NOMA and RAMA-I. To this end, we consider two extreme cases as follows. By definition, sum rate for NOMA and RAMA-I are $R_\text{sum}^{N} = R_1^N + R_2^N$ and $R_\text{sum}^{R,I} = R_1^{R,I} + R_2^{R,I}$, respectively, where $R_1^N$ and $R_2^N$ are defined in (\ref{eq33}) and $R_1^{R,I}$ and $R_2^{R,I}$ are defined in (\ref{eq7}).   

\subsubsection*{Case I} $p|h_1|^2/\sigma^2_1 = p|h_2|^2/\sigma^2_2$. In his paper is case is called symmetric channel~\cite{higuchi2015non}. That is, the two users have the same SNR. In this case, RAMA-I always achieves higher sum rate than NOMA.  

To show this, we calculate the sum rate for NOMA and RAMA. For NOMA, the sum rate can be calculated as 
\begin{align} \label{proof1}
R_\text{sum}^N &= \text{log}_2\big(1 + \frac{p_1|h_1|^2}{\sigma_1^2}\big) + \text{log}_2\big(1 + \frac{p_2|h_2|^2/\sigma_2^2}{p_1|h_2|^2/\sigma_2^2 + 1}\big) \nonumber \\
& \overset{(a)}{=} \text{log}_2\bigg(\big(1 + \frac{p_1|h_1|^2}{\sigma_1^2}\big)\big(1 + \frac{p_2|h_2|^2/\sigma_2^2}{p_1|h_2|^2/\sigma_2^2 + 1}\big)\bigg) \nonumber \\
& = \text{log}_2\big(1 + \frac{p_1|h_1|^2}{\sigma_1^2} + \frac{p_2|h_2|^2}{\sigma_2^2}\big) \nonumber \\
&\overset{(b)}{=} \text{log}_2\big(1 + \frac{p|h|^2}{\sigma^2}\big).
\end{align}
The $(a)$ follows from $\text{log}_2a + \text{log}_2b = \text{log}_2ab$ and the $(b)$ follows the assumption that $|h|^2/\sigma^2 = |h_1|^2/\sigma^2_1 = |h_2|^2/\sigma^2_2$ and $p_1 + p_2 = p$. 

Also, for RAMA-I, we follow the same steps as in NOMA. Hence, it is obtained as
\begin{equation}\label{proof2}
R_\text{sum}^{R,I} = \text{log}_2\big(1 + \frac{p|h|^2}{\sigma^2} + \frac{p^2|h|^4}{4\sigma^4}\big).
\end{equation} 
Since $ p^2|h|^4/4\sigma^4 > 0$, it gives $R_\text{sum}^N \leq R_\text{sum}^{R,I}$. 
 
\subsubsection*{Case II}  $p|h_1|^2/\sigma^2_1 \geq p|h_2|^2/\sigma^2_2$. Here, this case is called asymmetric channel~\cite{higuchi2015non}. That is, we assume that the channel gain for User 1 is stronger than User 2. It can be shown that for asymmetric channels, RAMA-I achieves higher sum rate than NOMA when the power isproperly  allocated  for Users 1 and 2\footnote{To achieve user fairness in NOMA, when $|h_1|^2/\sigma_1^2 \geq |h_2|^2/\sigma_2^2$ we have $p_1 \leq p_2$~\cite{higuchi2015non}.}. To proof this claim, we have 
\begin{align}\label{eq7a}
R_\text{sum}^N & = \text{log}_2(1 + \frac{p_1|h_1|^2}{\sigma^2_1}) + \text{log}_2(1 + \frac{p_2|h_2|^2/\sigma^2_2}{p_1|h_2|^2/\sigma^2_2 + 1}) \nonumber \\
& \overset{(a)}{=} \text{log}_2\big((1 + \frac{p_1|h_1|^2}{\sigma^2_1})(1 + \frac{p_2|h_2|^2/\sigma_2^2}{p_1|h_2|^2/\sigma_2^2 + 1})\big) \nonumber \\
& \overset{(b)}{\leq}  \text{log}_2\big((1 + \frac{p_1|h_1|^2}{\sigma^2_1})(1 + \frac{p_2|h_2|^2}{ \sigma^2_2})\big),
\end{align}
where $p_1$ and $p_2$ are allocated power for Users 1 and  2, respectively. Also, the (a) follows from $\text{log}_2a + \text{log}_2b = \text{log}_2ab$, and the (b) is a result of $p_1|h_2|^2/\sigma^2_2 > 0$ . Using (\ref{eq7a}) and (\ref{eq7}), the inequality $R^N_\text{sum} \leq R^{R,I}_\text{sum}$ holds when the following condition follows
\begin{align} \label{eq7b}
(1 + \frac{p_1|h_1|^2}{\sigma_1^2})(1 + \frac{p_2|h_2|^2}{ \sigma^2_2}) &\leq (1 + \frac{p|h_1|^2}{2\sigma_1^2})(1 + \frac{p|h_2|^2}{2 \sigma_2^2}).
\end{align}
Obviously, for $p_1/p \in (0,0.5]$ the inequality holds which indicates that User 1 should have lower power than User 2. 
Although this range is not tight, it gives a considerable insight. This result implies that with proper power allocation in NOMA, RAMA-I attains higher sum rate. However, RAMA-I may not achieve user fairness when channel gain of one of the users is significantly greater than that of other user. In this case, the allocated power for user with strong channel gain should be far less than other user and equal power allocation would not lead to user fairness.

\subsection{RAMA with full CSI}
\begin{figure}
\includegraphics[scale = 0.25]{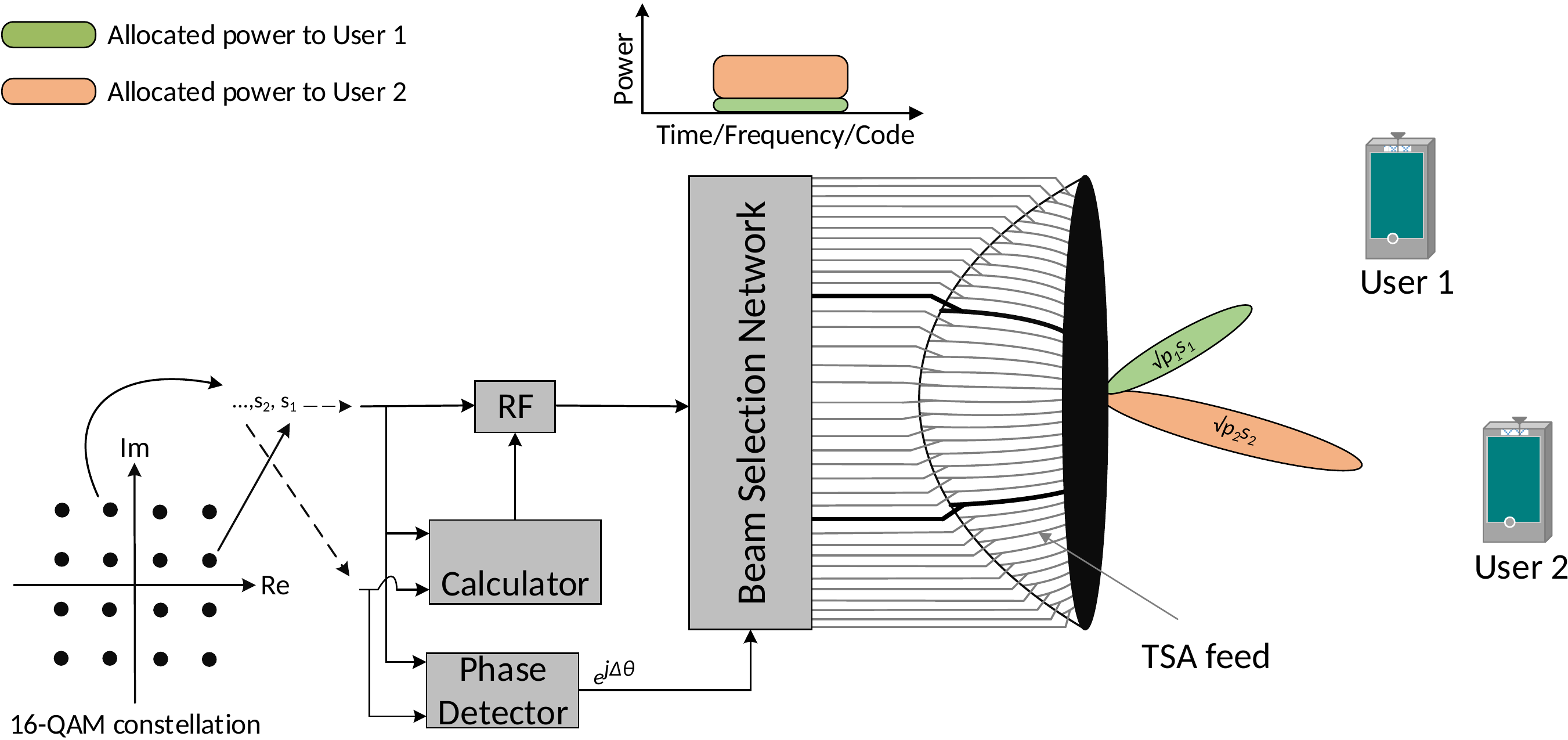}
\centering
\caption{Schematic of the BS for reconfigurable antenna multiple access technique regarding full CSI and unequal power allocation. It is assumed that signals $s_1$ and $s_2$ are chosen from 16-QAM constellation.}
 \label{fig3}
\end{figure}

Assume that full CSI is available at the BS. Furthermore, the BS can unequally allocate the power between two users. For signals $s_1$ and $s_2$ that are chosen from a QAM constellation and $\mathbb{E}[|s_i|^2] = 1$, the relationship between two arbitrary signals selected from the constellation is given by   
\begin{align}\label{eq8}
s_2 = s_1 \bar{s} e^{j\Delta \theta}
\end{align}
where $\bar{s}$ denotes $|s_2|/|s_1|$. For RAMA with full CSI, the transmitted signal is defined as
\begin{align}\label{eq8a}
x = \sqrt{p^\prime}s_1 \equiv (\sqrt{p_1 + p_2\bar{s}^2})s_1,
\end{align}
which obviously has the same average power as the signal in (\ref{eq1}). It is assumed that $p_1$ and $p_2$ are the allocated power for User 1 and User 2, respectively. For simplicity, we consider that our power allocation strategy is exactly the same as NOMA. Fig.~\ref{fig3} depicts the schematic of the reconfigurable antenna for the RAMA. The phase detector and $\bar{s}$ calculator blocks calculate $e^{j\Delta\theta}$ and $\bar{s}$, respectively. The beam selection network first selects two suitable TSA feeds, TSA 1 and TSA 2 which are highlighted with black color in Fig.~\ref{fig3}, based on the users' DoA information. Then, it divides the signal in (\ref{eq8a}) into $\sqrt{p_1}s_1$ and $\sqrt{p_2}s_1\bar{s}$ regarding the obtained CSI and the power allocation strategy. The signal for User 1, $\sqrt{p_1}s_1$, is  ready to transmit from TSA 1. For User 2, the intended signal is built by multiplying $\sqrt{p_2}s_1\bar{s}$ by $e^{j\Delta\theta}$ which yields $\sqrt{p_2}s_2$. Hence, the interference-free received signals for Users 1 and 2 are attained as     
\begin{align}\label{eq9}
\begin{cases}
z_1 = \sqrt{p_1}s_1 + n_1,\\
z_2 = \sqrt{p_2}s_2 + n_2,\\
\end{cases}
\end{align}
respectively. Accordingly, the achievable rate for user 1 and user 2 is obtained as
\begin{align}\label{eq10}
\begin{cases}
R^{R,\textit{II}}_1 = \text{log}_2(1 + \frac{p_1|h_1|^2}{\sigma^2_1}), \\
R^{R,\textit{II}}_2 = \text{log}_2(1 + \frac{p_2|h_2|^2}{\sigma^2_2}), 
\end{cases}
\end{align} 
respectively.

It is straightforward to show that RAMA with full CSI, denoted by RAMA-II, achieves higher sum rate than NOMA irrespective of their channel condition. That is, $R_\text{sum}^N \leq R_\text{sum}^{R,\textit{II}}$ where $R_\text{sum}^{R,\textit{II}} = R_1^{R,\textit{II}} + R_2^{R,\textit{II}}$. Further, RAMA-II considers user fairness the same as NOMA.       

\subsection*{The following remarks are in order:} 
\begin{itemize}
\item RAMA-I significantly reduces system overhead. In highly dense networks where the number of users is much larger than two, it is reasonable to serve users with RAMA-I. This is because RAMA-I does not require the BS to know full CSI and only users' direction information is necessary. 
\item RAMA-I implemented by PSK constellation is very simple in practice. The beam selection network divides the power equally by using a simple power divider and the received signal is interference-free. 
The later is also preserved for RAMA-II. Indeed, RAMA-I with QAM constellations is operational if power is divided properly.  
\item RAMA-II needs an optimal power allocation strategy. In Subsection III.B it is pointed out that power allocation strategy for NOMA is adopted for RAMA-II. However, this strategy may not be efficient in reconfigurable antenna systems with full CSI. It is because NOMA considers the interference and the minimum achievable rate for each user to allocate the power~\cite{higuchi2015non}. Whereas, for RAMA-II the interference is removed which leads to designing a better power allocation strategy.   
\item RAMA technique is not an alternative for NOMA. When users are aligned in the same direction, RAMA would be combined with one of OMA or NOMA. Accordingly, the users located on the same direction are considered as a cluster. Each cluster will be served via RAMA-OMA or RAMA-NOMA. Integration of RAMA with other multiple access techniques is beyond the scope of this paper.   
\end{itemize}
\section{Numerical Computations}
This section evaluates the performance of the proposed multiple access technique by using numerical computations, where the analytical findings will also be verified. Transmission in mmWave bands can be done through both line-of-sight and non line-of-sight paths. Here, for the sake of simplicity, we will consider Rayleigh fading channels.

\begin{figure}[b]
    \centering
    \subfloat[ ]{{\includegraphics[scale=0.4]{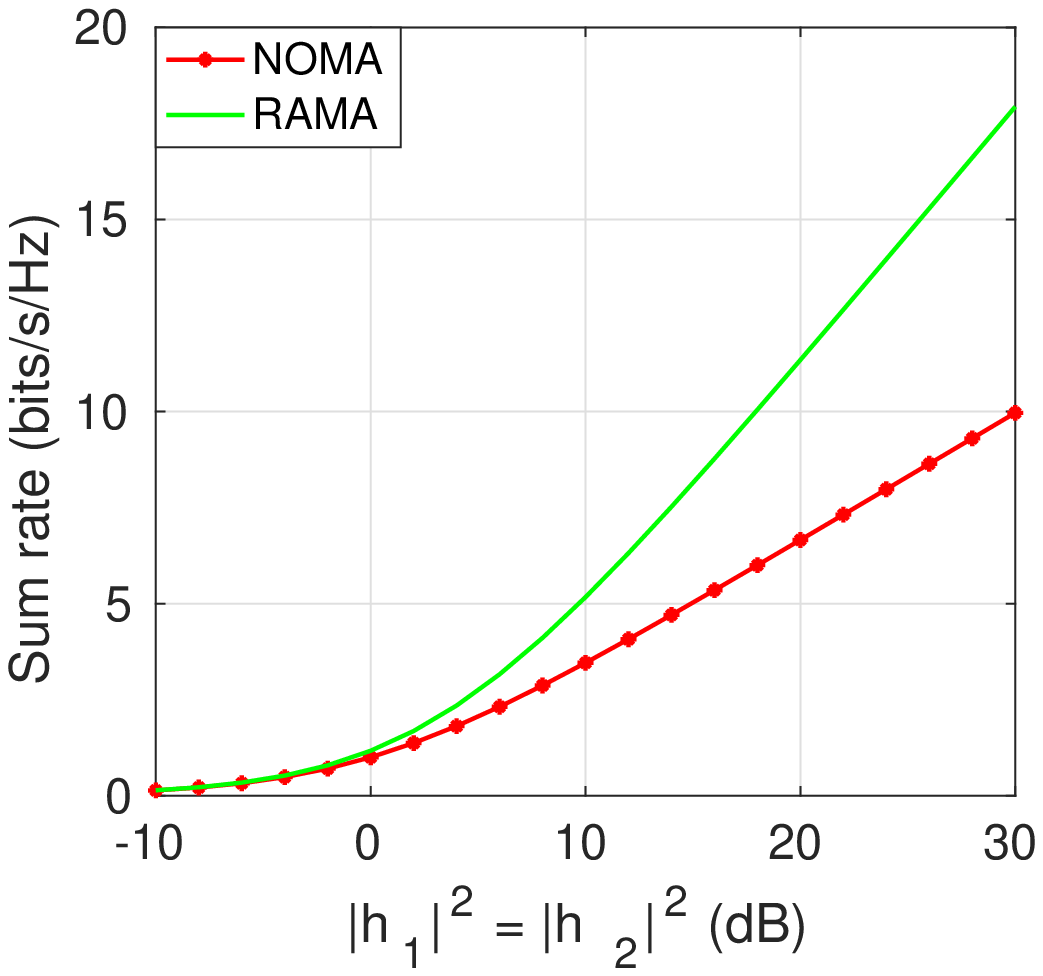} }}
    \subfloat[ ]{{\includegraphics[scale=0.4]{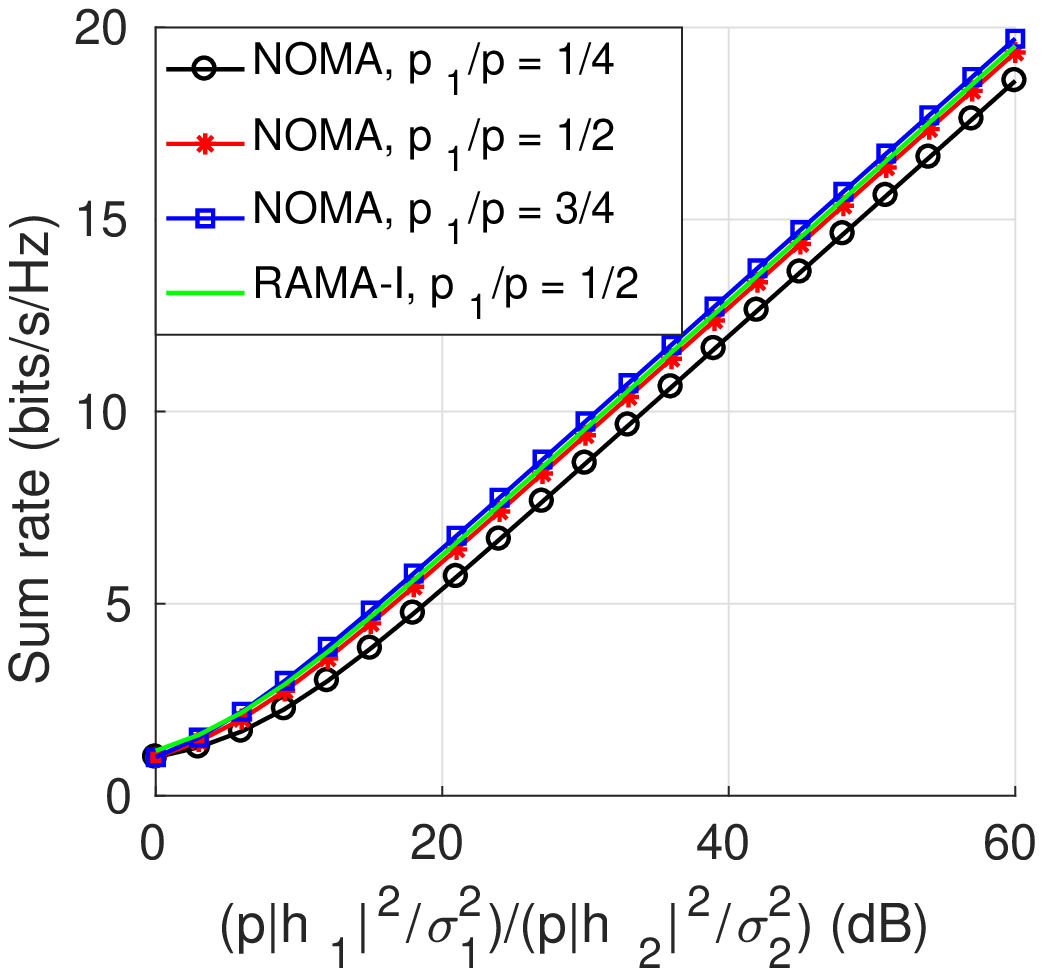} }}
    \caption{Sum rate comparison between NOMA and RAMA-I for (a) symmetric channel and (b) symmetric channel.}
    \label{fig45}
\end{figure}

Figure~\ref{fig45}.(a) represents sum rate versus symmetric channel plot for NOMA and RAMA-I. It is clear that RAMA-I achieves better sum rate than NOMA. This is because the user achievable rate for NOMA is limited due to the interference from other users, which degrades the sum rate. At symmetric channels, the interference in NOMA is severe and as a result leads to a considerable sum rate gap compared to the RAMA technique which is an inter-user interference-free technique. This result verifies our claim in \textit{Case I} Subsection~III.A.

Figure~\ref{fig45}.(b) illustrates sum rate performance versus asymmetric channel gains for RAMA-I and various power allocation schemes for NOMA. The aim of this simulation is to support our analytical finding in \text{Case II} Subsection~III.A. When $(p|h_1|^2/\sigma^2_1)/(p|h_2|^2/\sigma^2_2)$ is not large enough, RAMA-I outperforms NOMA for all power allocation schemes since the channel is similar to a symmetric channel. By increasing $(p|h_1|^2/\sigma^2_1)/(p|h_2|^2/\sigma^2_2)$ channel satisfies the condition in $\textit{Case II}$ Subsection~III.A. At high region of $(p|h_1|^2/\sigma^2_1)/(p|h_2|^2/\sigma^2_2)$, for $p_1/p \leq 1/2$, e.g., $p_1/p = 1/2$ and $p_1/p = 1/4$, sum rate of RAMA-I is always better than NOMA which is consistent with \textit{Case II}. For $p_1/p = 3/4$ and large channel gain difference, NOMA has a little better sum rate. This is because much more power is allocated to User 1 which can nearly achieve maximum sum rate. However, in this condition NOMA does not consider user fairness. In contrast, equal power is allocated for the users in RAMA-I and they cannot exploit maximum sum rate.

\begin{figure}
    \centering
    \subfloat[ ]{{\includegraphics[scale=0.4]{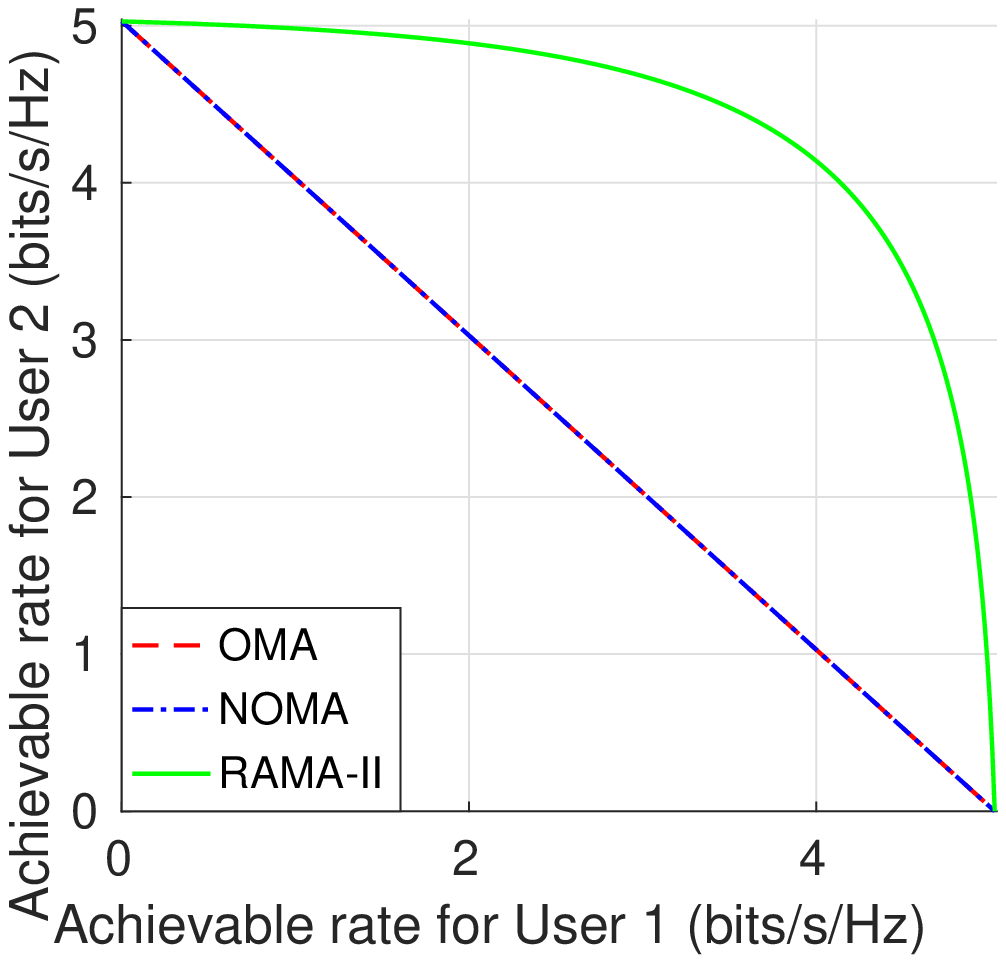} }}
    \subfloat[ ]{{\includegraphics[scale=0.4]{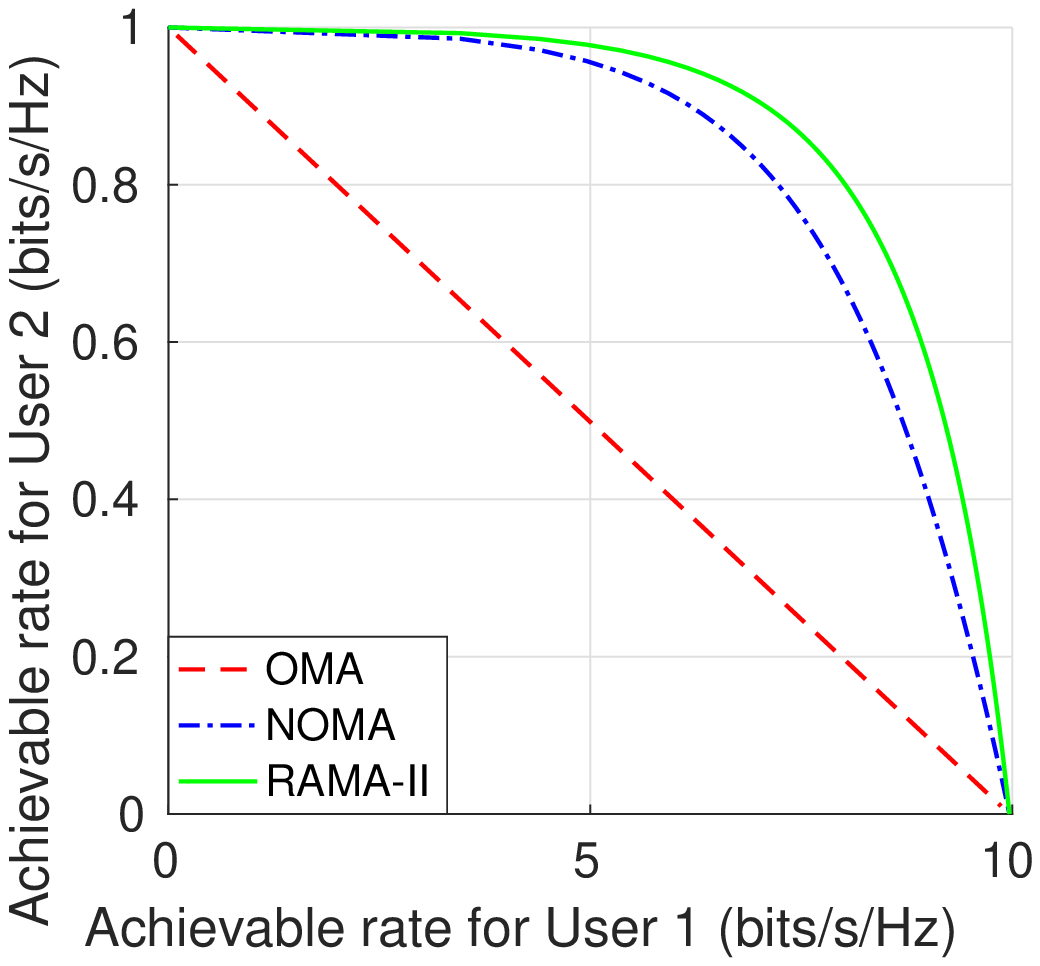} }}
    \caption{Achievable rate region of user 1 and 2 for (a) symmetric channel with $p|h_i|^2/\sigma_i^2 = 15$ dB for $i = 1, 2$ and (b) asymmetric channel with $p|h_1|^2/\sigma_1^2 = 30$ dB and $p|h_2|^2/\sigma_2^2 = 0$ dB.}
    \label{fig67}
\end{figure}

Figure~\ref{fig67} shows achievable rate region of two users for OMA, NOMA, and RAMA-II. OMA in downlink transmission is assumed to be implemented by OFDMA technique where $R_1^O = \beta \text{log}_2(1 + p_1|h_1|^2/\beta\sigma_1^2)$ and $R_2^O = (1 - \beta) \text{log}_2(1 + p_2|h_2|^2/(1 - \beta)\sigma_2^2)$ are achievable rate for Users 1 and 2 with the bandwidth of $\beta$ Hz assigned to User 1 and $(1 - \beta)$ Hz assigned to User 2~\cite{higuchi2015non}.  

In Fig.~\ref{fig67}.(a), channel is assumed to be symmetric where its gain is set to $p|h_i|^2/\sigma_i^2 = 15$ dB for $i = 1, 2$. The achievable rate region for OMA and NOMA are identical. The region for RAMA-II is much wider than that for OMA and NOMA because RAMA-II neither suffers from inter-user interference nor divides the bandwidth among users. For instance, when User 2 achieves rate 2.5 bits/s/Hz, achievable rate of User 1 for RAMA-II with channel gain information is approximately twice higher than that for OMA and NOMA. Notice that when total power is allocated to either user, all three techniques are able to achieve maximum rate for that user.         

The achievable sum rate of asymmetric channel for $p|h_1|^2/\sigma_1^2 = 30$ dB and $p|h_2|^2/\sigma_2^2 = 0$ dB has been represented in Fig.~\ref{fig67}.(b). From the figure it is clear that for OMA, NOMA and RAMA-II only User 1 achieves maximum sum rate when whole power is allocated to that user. However, NOMA achieves wider rate region than OMA as expected. Interestingly, the achievable rate region for RAMA-II is greater than NOMA. As an example, when we want that User 1 to achieve 8 bits/s/Hz, User 1 can reach 0.68 bits/s/Hz and 0.8 bits/s/Hz with NOMA and RAMA-II, respectively. This is because the achievable rate of User 2 in NOMA is affected by the interference term from User 1 due to using superposition coding at  transmitter and SIC process at receiver. Whereas, in RAMA-II, User 1 does not impart interference on User 2. In other words, the gap between NOMA and RAMA-II reflects the impact of inter-user interference on NOMA. 
\section{Conclusion}
In this paper, we propose a new multiple access technique for mmWave reconfigurable antennas in order to simultaneously support two users by using a single BS in downlink. First, we show that NOMA is not suitable technique for serving the users for the reconfigurable antenna systems. Then, by wisely using the properties of reconfigurable antennas, a novel multiple access technique called RAMA is designed by assuming partial CSI and full CSI. The proposed RAMA provides mmWave reconfigurable antenna system with an inter-user interference-free user serving. That is, the users with higher allocated power do not required to decode signals of other users. It is shown that for symmetric channels RAMA-I outperforms NOMA for an arbitrary $p_1$ in terms of sum rate. Also, for asymmetric channels RAMA-I demonstrates better sum rate performance if approximately more than half of the power is allocated to User 2. Further, RAMA-II always achieves higher sum rate than NOMA. Numerical computations support our analytical investigations.

\appendices

\bibliographystyle{IEEEtran}
\bibliography{IEEEabrv,references}

\begin{thebibliography}{10}
\providecommand{\url}[1]{#1}
\csname url@samestyle\endcsname
\providecommand{\newblock}{\relax}
\providecommand{\bibinfo}[2]{#2}
\providecommand{\BIBentrySTDinterwordspacing}{\spaceskip=0pt\relax}
\providecommand{\BIBentryALTinterwordstretchfactor}{4}
\providecommand{\BIBentryALTinterwordspacing}{\spaceskip=\fontdimen2\font plus
\BIBentryALTinterwordstretchfactor\fontdimen3\font minus
  \fontdimen4\font\relax}
\providecommand{\BIBforeignlanguage}[2]{{%
\expandafter\ifx\csname l@#1\endcsname\relax
\typeout{** WARNING: IEEEtran.bst: No hyphenation pattern has been}%
\typeout{** loaded for the language `#1'. Using the pattern for}%
\typeout{** the default language instead.}%
\else
\language=\csname l@#1\endcsname
\fi
#2}}
\providecommand{\BIBdecl}{\relax}
\BIBdecl

\bibitem{r1}
T.~S. Rappaport, S.~Sun, R.~Mayzus, H.~Zhao, Y.~Azar, K.~Wang, G.~N. Wong,
  J.~K. Schulz, M.~Samimi, and F.~Gutierrez, ``Millimeter wave mobile
  communications for 5{G} cellular: {I}t will work!'' \emph{IEEE Access},
  vol.~1, pp. 335--349, 2013.

\bibitem{r5}
J.~Kim and I.~Lee, ``802.11 {WLAN}: history and new enabling {MIMO} techniques
  for next generation standards,'' \emph{{IEEE} Commun. Mag.}, vol.~53, no.~3,
  pp. 134--140, 2015.

\bibitem{el2014spatially}
O.~El~Ayach, S.~Rajagopal, S.~Abu-Surra, Z.~Pi, and R.~W. Heath, ``Spatially
  sparse precoding in millimeter wave mimo systems,'' \emph{{IEEE} Trans.
  Wireless Commun.}, vol.~13, no.~3, pp. 1499--1513, 2014.

\bibitem{r9}
J.~Brady, N.~Behdad, and A.~M. Sayeed, ``Beamspace {MIMO} for millimeter-wave
  communications: {S}ystem architecture, modeling, analysis, and
  measurements,'' \emph{{IEEE} Trans. Antennas Propagat.}, vol.~61, no.~7, pp.
  3814--3827, 2013.

\bibitem{higuchi2015non}
K.~Higuchi and A.~Benjebbour, ``Non-orthogonal multiple access ({NOMA}) with
  successive interference cancellation for future radio access,'' \emph{IEICE
  Trans. Commun.}, vol.~98, no.~3, pp. 403--414, 2015.

\bibitem{dai2015non}
L.~Dai, B.~Wang, Y.~Yuan, S.~Han, I.~Chih-Lin, and Z.~Wang, ``Non-orthogonal
  multiple access for 5{G}: solutions, challenges, opportunities, and future
  research trends,'' \emph{{IEEE} Commun. Mag.}, vol.~53, no.~9, pp. 74--81,
  2015.

\bibitem{tse2005fundamentals}
D.~Tse and P.~Viswanath, \emph{Fundamentals of wireless communication}.\hskip
  1em plus 0.5em minus 0.4em\relax Cambridge university press, 2005.

\bibitem{saito2013system}
Y.~Saito, A.~Benjebbour, Y.~Kishiyama, and T.~Nakamura, ``System-level
  performance evaluation of downlink non-orthogonal multiple access ({NOMA}),''
  in \emph{Proc. IEEE Int. Symp. Pers., Indoor Mobile Radio Commun.
  (PIMRC)}.\hskip 1em plus 0.5em minus 0.4em\relax IEEE, Sep. 2013, pp.
  611--615.

\bibitem{saito2013non}
Y.~Saito, Y.~Kishiyama, A.~Benjebbour, T.~Nakamura, A.~Li, and K.~Higuchi,
  ``Non-orthogonal multiple access ({NOMA}) for cellular future radio access,''
  in \emph{Proc. IEEE VTC Spring}.\hskip 1em plus 0.5em minus 0.4em\relax IEEE,
  2013, pp. 1--5.

\bibitem{ding2017random}
Z.~Ding, P.~Fan, and H.~V. Poor, ``Random beamforming in millimeter-wave {NOMA}
  networks,'' \emph{IEEE Access}, 2017.

\bibitem{ding2017noma}
Z.~Ding, L.~Dai, R.~Schober, and H.~V. Poor, ``{NOMA} meets finite resolution
  analog beamforming in massive {MIMO} and millimeter-wave networks,''
  \emph{{IEEE} Commun. Lett.}, 2017.

\bibitem{wang2017spectrum}
B.~Wang, L.~Dai, Z.~Wang, N.~Ge, and S.~Zhou, ``Spectrum and energy-efficient
  beamspace {MIMO-NOMA} for millimeter-wave communications using lens antenna
  array,'' \emph{{IEEE} J. Select. Areas Commun.}, vol.~35, no.~10, pp.
  2370--2382, 2017.

\bibitem{hao2017energy}
W.~Hao, M.~Zeng, Z.~Chu, and S.~Yang, ``Energy-efficient power allocation in
  millimeter wave massive {MIMO} with non-orthogonal multiple access,''
  \emph{IEEE Wireless Commun. Lett.}, vol.~6, no.~6, pp. 782--785, 2017.

\bibitem{xiao2017joint}
Z.~Xiao, L.~Zhu, J.~Choi, P.~Xia, and X.-G. Xia, ``Joint power allocation and
  beamforming for non-orthogonal multiple access ({NOMA}) in 5{G}
  millimeter-wave communications,'' \emph{arXiv preprint arXiv:1711.01380},
  2017.

\bibitem{r19}
M.~A. Almasi, H.~Mehrpouyan, V.~Vakilian, N.~Behdad, and H.~Jafarkhani,
  ``Reconfigurable antennas in mm{W}ave {MIMO} systems,'' \emph{arXiv preprint
  arXiv:1710.05111}, 2017.

\bibitem{schoenlinner2002wide}
B.~Schoenlinner, X.~Wu, J.~P. Ebling, G.~V. Eleftheriades, and G.~M. Rebeiz,
  ``Wide-scan spherical-lens antennas for automotive radars,'' \emph{{IEEE}
  Trans. Microwave Theory Tech.}, vol.~50, no.~9, pp. 2166--2175, 2002.

\end{thebibliography}
\end{document}